\documentclass[fleqn,10pt]{wlscirep}
\title{Small-scale displacement fluctuations of vesicles in fibroblasts}

\author[1]{Danielle Posey}
\author[2]{Paris Blaisdell-Pijuan}
\author[3]{Samantha K. Knoll}
\author[3]{Taher A. Saif}
\author[2,*]{Wylie W. Ahmed}
\affil[1]{Department of Biological Sciences, California State University, Fullerton, CA}
\affil[2]{Department of Physics, California State University, Fullerton, CA}
\affil[3]{Department of Mechanical Science and Engineering, University of Illinois, Urbana, IL}

\affil[*]{Corresponding author (wahmed@fullerton.edu)}

\begin{abstract}
The intracellular environment is a dynamic space filled with various organelles moving in all directions.  Included in this diverse group of organelles are vesicles, which are involved in transport of molecular cargo throughout the cell.  Vesicles move in either a directed or non-directed fashion, often depending on interactions with cytoskeletal proteins such as microtubules, actin filaments, and molecular motors.  How these proteins affect the local fluctuations of vesicles in the cytoplasm is not clear since they have the potential to both facilitate and impede movement.  Here we show that vesicle mobility is significantly affected by myosin-II, even though it is not a cargo transport motor.  We find that myosin-II activity increases the effective diffusivity of vesicles and its inhibition facilitates longer states of non-directed motion. Our study suggests that altering myosin-II activity in the cytoplasm of cells can modulate the mobility of vesicles, providing a possible mechanism for cells to dynamically tune the cytoplasmic environment in space and time.

\end{abstract}
\begin{document}

\flushbottom
\maketitle

\thispagestyle{empty}

\section*{Introduction}

The cell interior is a dynamic environment where vesicles and other organelles traverse the cytoplasm to localize for specific biological processes or to evenly disperse throughout the cell \cite{ross2008cargo}.  Vesicle motion can be categorized into two main regimes: directed and non-directed \cite{Caspi_Diffusion_2002,Arcizet_Temporal_2008,Ahmed_Measuring_2012,koslover2017cytoplasmic}.  Directed intracellular transport is critical to maintain normal cell function and it is typically driven by molecular motors that consume energy and transport cargo along cytoskeletal filaments \cite{Hirokawa_The_2009,Roberts_Functions_2013,Schuh_An_2011}.  Non-directed vesicle motion is largely driven by thermal fluctuations and results in random diffusion in the complex cytoplasmic environment; although, recent studies show that it can be driven by molecular motors \cite{Fakhri_High_2014,Almonacid_Active_2015,Brangwynne_Intracellular_2009,Ahmed_ActiveOocyte} and possibly other athermal processes \cite{golestanian2015enhanced,sengupta2013enzyme} and is called active diffusion.  Observations of vesicle movement in cells reveal vesicles stochastically alternating between directed and non-directed motion, mixing the dynamics and leading to anomalous motion \cite{Hfling_Anomalous_2013,Goychuk_How_2014,Gal_Experimental_2010,Toli-Nrrelykke_Anomalous_2004,Salman_Microtubules_2002,Regner_Identifying_2014,Saxton_A_2007,Weiss_Anomalous_2004,Dix_Crowding_2008,Brangwynne_Intracellular_2009,Fakhri_High_2014}.  Sub-diffusive motion of vesicles in cells is widely observed and often attributed to the soft-glassy nature of the cytoskeleton \cite{fabry2001scaling}.

Several techniques have been developed to analyze this stochastic motion \cite{Arcizet_Temporal_2008,Ahmed_Measuring_2012,Chen_Diagnosing_2013,gal2013particle,koslover2017cytoplasmic}.  Most studies of vesicle movement focus on timescales of seconds, minutes, or even hours where long range transport occurs \cite{Schuh_An_2011,hancock2014bidirectional,rogers2000membrane}.  In this study, we focus on the small-scale displacement fluctuations of vesicle motion on timescales from milliseconds to seconds to provide insight on the local mobility of the cytoplasmic microenvironment. We note that we have not specifically labeled the vesicles tracked in this study and do not have information regarding their molecular identity.  We simply employ these vesicles as generic tracers that report on the local mobility in the cytoplasm.

Vesicle movement is an inherently mechanical process, where motion in the cytoplasm requires a force and the vesicle experiences a corresponding resistance (viscoelastic drag force) from the surrounding environment \cite{Camps_A_2016}.  In this picture, the cytoskeleton can simultaneously serve to facilitate mobility via molecular motor driven transport, and to hinder mobility since its dense meshwork could act as a barrier.  In either case, the cytoskeleton is expected to affect vesicle motion in the cytoplasm.  Further, cells exhibit different cytoskeletal structures and higher bulk stiffness on increasing substrate stiffness \cite{Solon2007fibstiff}.  To this end, we investigate two stimuli known to affect cytoskeletal mechanical properties: substrate stiffness and pharmacological drugs targeted to the cytoskeleton.  We used glass and polyacrylamide gels ($E = 40$ kPa and $10$ kPa) to study the effect of substrate stiffness.  To discover the roles of the major cytoskeletal components in vesicle motion we introduced the pharmacological agents cytochalasin-D (cyto-D), colchicine, and blebbistatin, to target and depolymerize actin filaments, microtubule filaments, and inactivate myosin-II, respectively.

We used high-speed differential interference contrast microscopy to record vesicle motion in cells and analyze the effect of substrate stiffness and pharmacological drugs.  Vesicle dynamics were quantified via mean squared displacement (MSD), velocity autocorrelation function (VACF), and van Hove correlation (VHC) analyses.  We find that vesicle mobility in normal cells is independent of substrate stiffness, however, when pharmacological treatments are used their effect increases on softer substrates.  Furthermore, our results suggest that small-scale directed vesicle motion is robust, and the actin-myosin-II network plays a role in modulating local vesicle diffusion in the cytoplasm.


\section*{Results}
\begin{figure}[ht]
\centering
\includegraphics[width=\linewidth]{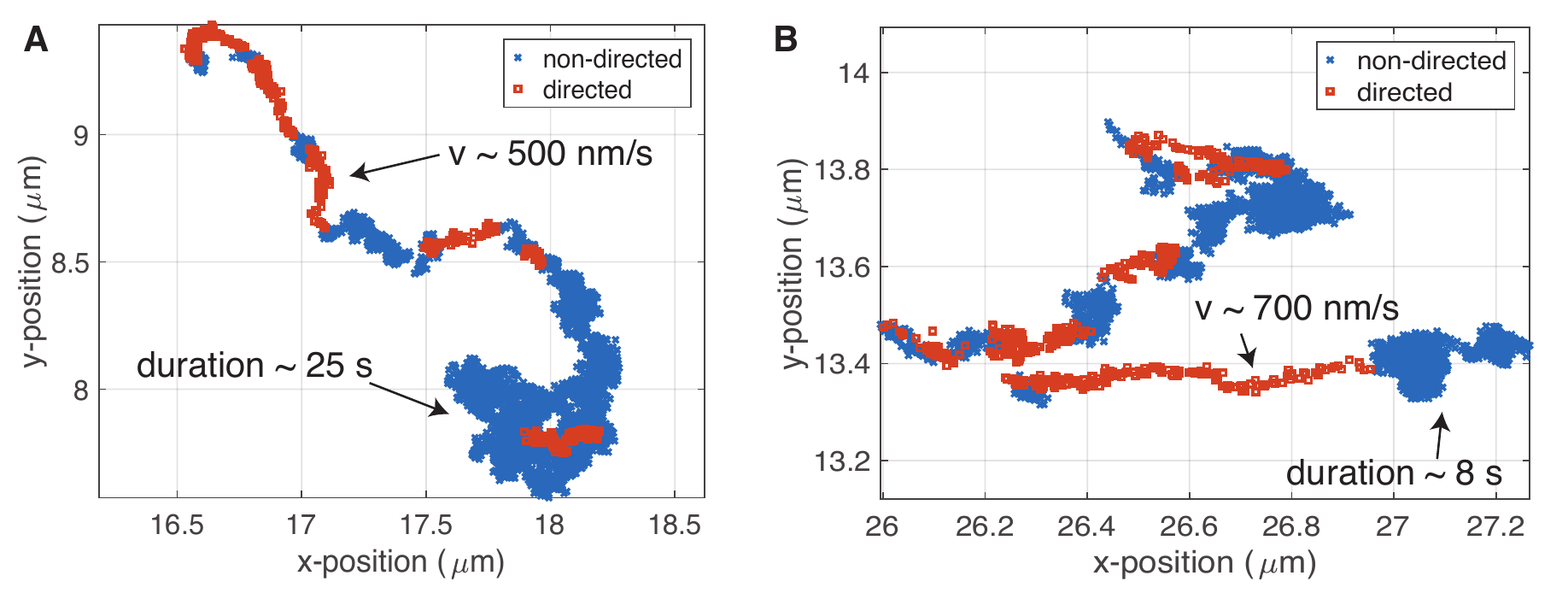}
\caption{\textbf{Trajectories of vesicle motion in fibroblasts.} Two examples of raw data for the vesicle positions are shown.  The temporal mean squared displacement ($t$MSD) analysis was used to identify directed and non-directed motion. (A) A trajectory exhibiting short bursts of directed motion ($v \sim 500$ nm/s) followed by a long period of non-directed motion  ($\sim$ 25 s). (B) Another trajectory showing  directed motion ($v \sim 700$ nm/s) and non-directed motion ($\sim$ 8 s).  The maximum possible trajectory duration in this study is 30 s due to acquisition length.}
\label{fig:t_MSDexample}
\end{figure}

All experiments were conducted using differential interference contrast microscopy to visualize vesicle motion in fibroblasts.  Vesicles were tracked using an algorithm based on polynomial fitting with Gaussian weight, specifically developed for tracking endogenous vesicles in the cell with nanometer precision using widefield microscopy \cite{rogers2007precise}.   To quantify small-scale vesicle displacements, the temporal MSD ($t$MSD) analysis is used as described previously \cite{Ahmed_Measuring_2012}.  In brief, the local MSD for each vesicle is calculated and fit to a power-law to determine the state of motion, where a power-law of 1 indicates non-directed motion, and a power-law of 2 indicates ballistic directed motion (see Materials and Methods).  The $t$MSD analysis allows determination of directed and non-directed motion in a single vesicle trajectory (Fig. \ref{fig:t_MSDexample}). Subsequently this allows the extraction of several parameters such as velocity and duration of each state of vesicle motion.  Throughout the study, directed and non-directed motion was determined via the $t$MSD analysis.

\subsection*{Vesicle motion is independent of substrate stiffness}
\begin{figure}[ht]
\centering
\includegraphics[width=\linewidth]{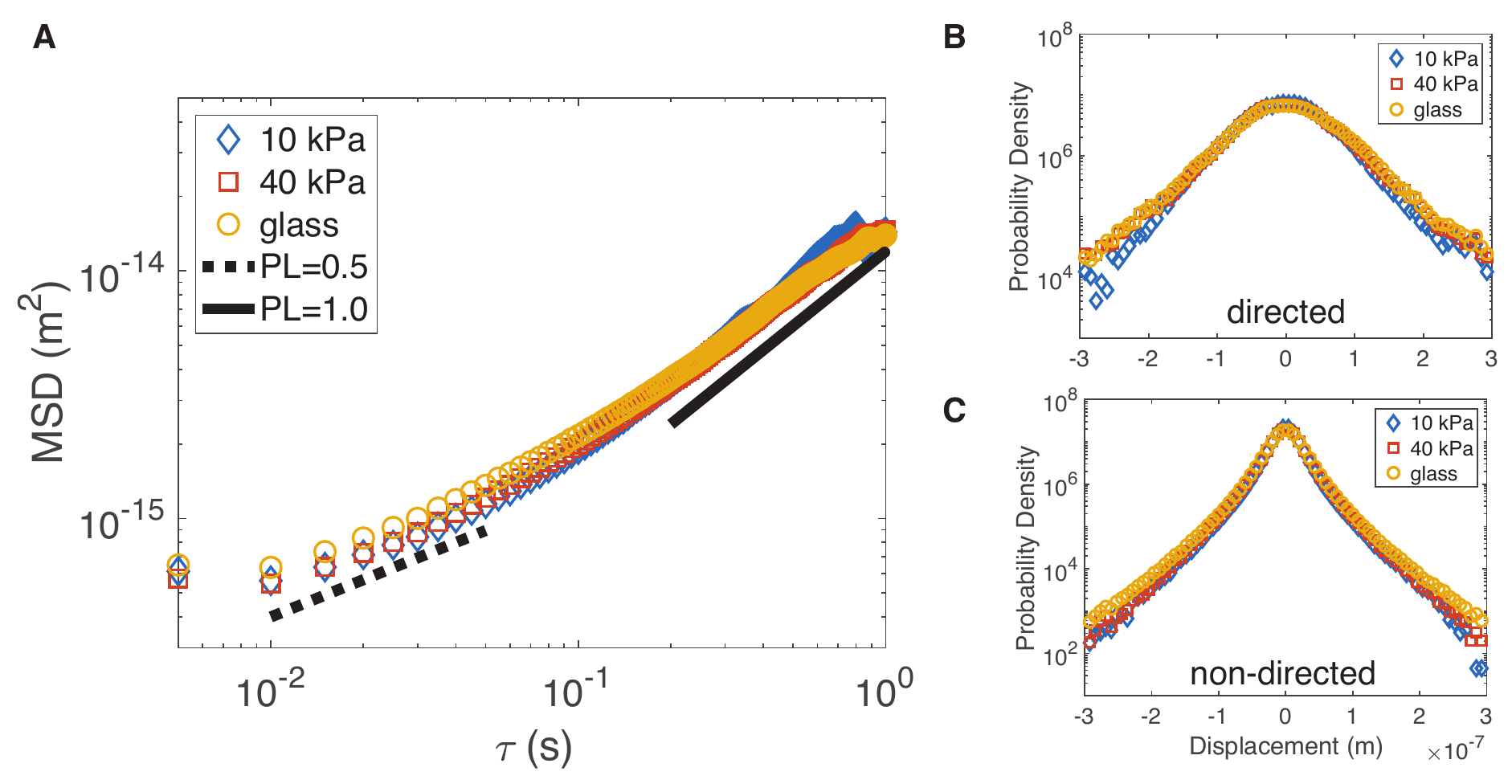}
\caption{\textbf{Vesicle dynamics does not depend on substrate stiffness.} (A) Mean squared displacement (MSD) of vesicle motion in untreated cells indicates no strong dependence on substrate stiffness.  Vesicles in all cases show sub-diffusive behavior at shorter timescales transitioning to diffusive behavior at longer timescales.  van Hove Correlation (VHC) functions for vesicles undergoing directed motion (B) and non-directed motion (C) show the full distribution of vesicle displacements is not sensitive to substrate stiffness.  VHCs calculated for a timescale of 150 ms.  Notice that VHCs of both directed and non-directed motion are strongly non-Gaussian, indicating cytoplasmic activity \cite{Ahmed_SciRep_2015}. (Number of experiments/cells/trajectories - 10 kPa: 7/28/1211, 40 kPa: 7/24/3021, glass: 10/31/5605; S.E.M. smaller than symbol size)}
\label{fig:MSD_stiffness}
\end{figure}

Control cells were cultured on substrates of varying rigidity ($E = 10$ kPa, 40 kPa, and glass).  Small-scale displacement fluctuations of vesicles in untreated fibroblasts were tracked and analyzed to characterize their overall motion on the three substrates.  The MSD was calculated for each substrate and showed that the small-scale fluctuations of vesicles were not strongly affected by substrate stiffness (Fig. \ref{fig:MSD_stiffness}A).  On all three surfaces, vesicle motion is sub-diffusive at short timescales and transitions to diffusive-like behavior at longer timescales.  Using the $t$MSD analysis, vesicle trajectories were split into directed and non-directed states to examine the full displacement distributions, also known as VHC functions \cite{Ahmed_SciRep_2015}.  Fig. \ref{fig:MSD_stiffness}B,C indicates that the displacement distribution of both directed and non-directed vesicle motions are not strongly affected by substrate stiffness.  Furthermore, the velocity distribution of vesicles undergoing directed motion also does not depend on the substrate (Fig. \ref{fig:VACF}A).  This result is unexpected, since substrate stiffness is known to affect cytoplasmic mechanics \cite{Solon2007fibstiff,Lo_Cell_2000}, however, overall vesicle mobility within the cytoplasm remains unperturbed.

\subsection*{Myosin-II inhibition decreases effective diffusivity of vesicles}
\begin{figure}[ht]
\centering
\includegraphics[width=\linewidth]{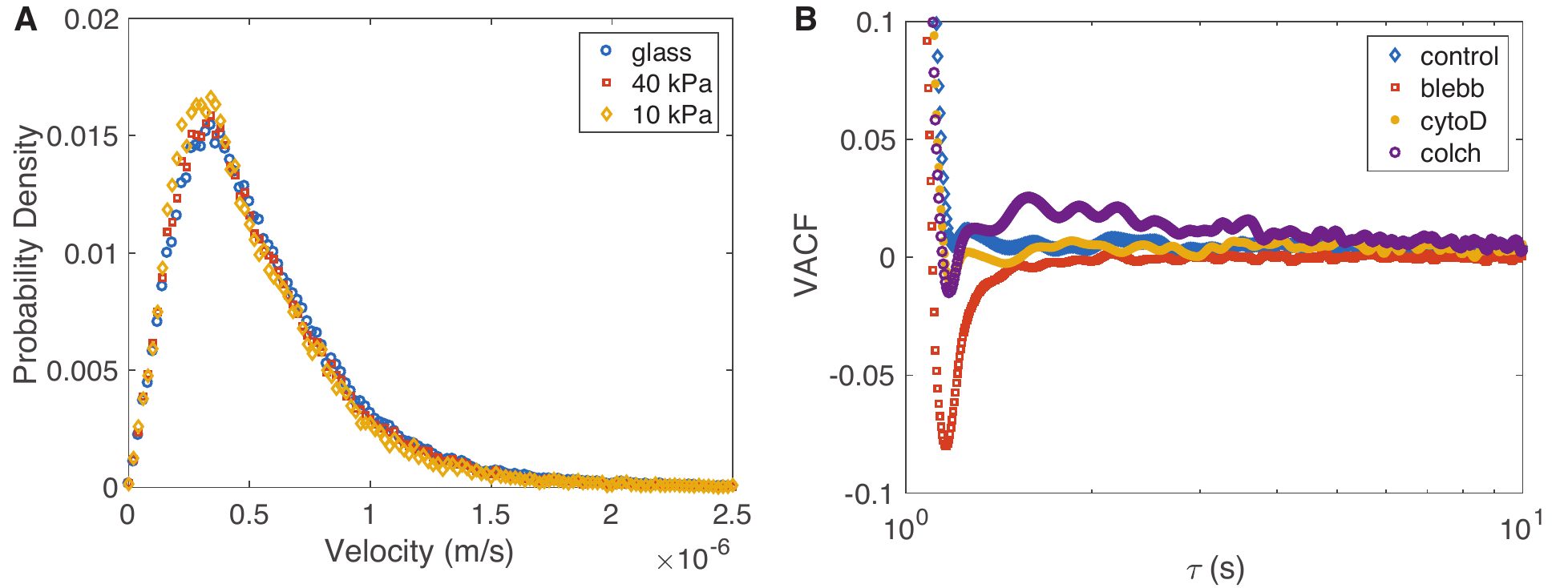}
\caption{\textbf{Velocity statistics indicate robustness to substrate stiffness and sensitivity to cytoskeletal perturbation.} (A) The velocity distribution of directed motion is independent of  substrate stiffness.  This provides further support, in addition to MSDs and VHCs, that substrate stiffness does not affect vesicle motion in fibroblasts. (B) The velocity autocorrelation function (VACF) quantifies persistence and indicates that vesicle motion is strongly perturbed by blebbistatin, which inactivates myosin-II.  The strong negative peak (red squares) indicates vesicle motion is decreased when myosin-II is inactivated. Data shown for cells cultured on 10 kPa substrate. }
\label{fig:VACF}
\end{figure}

To quantify the diffusivity of vesicles, the VACF was calculated and the effective diffusivity ($D_\mathrm{eff}$) was extracted via the Green-kubo relation (see Materials an Methods).  When cells are treated with blebbistatin, the $D_\mathrm{eff}$ of vesicles decreases by nearly 40-fold for cells on 10 kPa gels as evident by the blebbistatin data in Fig. \ref{fig:VACF}B and Table \ref{tab:Diffusion_coefficients}.  On stiffer surfaces (40 kPa and glass) the same effect is observed, however $D_\mathrm{eff}$ is decreased by 10-fold (Table \ref{tab:Diffusion_coefficients}).  Thus, in the absence of myosin-II activity, small-scale vesicle motion is greatly decreased.  The effect of perturbing the cytoskeleton itself seems to have less of an effect; depolymerizing actin (via cyto-D) or microtubules (via colchicine) does not significantly affect vesicle diffusivity, as $D_\mathrm{eff}$ remains within a factor of 2.  Overall these results (Table \ref{tab:Diffusion_coefficients}) suggest that myosin-II plays a significant role in maintaining vesicle diffusivity for small-scale motion.


\begin{table}[ht]
\centering
\begin{tabular}{|l|l|l|l|l|}
\hline
\textbf{Substrate Stiffness} & \textbf{Control} ($\mu$$m^{2}$/s) & \textbf{Blebbistatin} ($\mu$$m^{2}$/s) & \textbf{Cyto-D} ($\mu$$m^{2}$/s) & \textbf{Colchicine} ($\mu$$m^{2}$/s) \\
\hline
10 kPa & 1.80E-3 & 4.34E-5 & 8.12E-4 & 1.49E-3 \\
\hline
40 kPa & 1.90E-3 & 2.10E-4 & 8.69E-4 & 1.44E-3 \\
\hline
Glass & 1.54E-3 & 2.02E-4 & 7.37E-4 & 1.58E-3 \\
\hline
\end{tabular}
\caption{\label{tab:example} \textbf{Effective diffusion coefficients } ($D_\mathrm{eff}$ in $\mu$$m^{2}$/s)  Diffusion of vesicles is strongly affected by blebbistatin, which decreases $D_\mathrm{eff}$ by a factor of $\sim$10-40 depending on substrate stiffness. In control cells, as well as cyto-D and colchicine treated, diffusion is not strongly affected.}
\label{tab:Diffusion_coefficients}
\end{table}

\subsection*{Directed motion of vesicles is robust}
To take a closer look at the local dynamics of vesicles in the cytoplasm, we used a $t$MSD analysis to analyze the statistics (e.g. velocity and run-time) of directed and non-directed vesicle motion.  The run-time of a particular state (directed or non-directed) is the amount of time a vesicle remains in that state of motion as determined by the $t$MSD analysis \cite{Ahmed_Measuring_2012}.  We found that vesicles typically undergo small-scale directed motions with average velocities of $\sim 550$ nm/s and run-times of less than 500 ms (100 video frames) (Table \ref{tab:Mean_velocities}, \ref{tab:Directed_runtimes}).  The histograms of directed run-times were normalized to create a probability distribution function (Fig. \ref{fig:runtimes}A).  Thus, integrating over specific time intervals provides the probability of that event occurring.

To quantify the statistics of motion, we integrate the probability density functions.  Integrating the probability density function of directed run-times shows 94\% of vesicles in control, cyto-D, and colchicine treated cells experience directed motion of duration less than 500 ms, and in the case of blebbistatin-treated cells the number was slightly larger, 100\% (Fig. \ref{fig:runtimes}A, light blue shaded region and Table \ref{tab:Directed_runtimes}).  This result suggests the possibility that local directed transport of vesicles is not directly dependent on myosin-II, actin filaments, or microtubules.  Alternatively, it is possible that a subpopulation of stable actin and microtubules still remain after drug treatment and facilitate local directed motion.  It is important to note that this result applies to small-scale directed motion only, which may be strongly dependent on local cytoplasmic geometry (e.g. mesh size).  Long-range directed transport must undoubtedly depend on the presence of cytoskeletal filaments.

\subsection*{Myosin-II inhibition increases the duration of non-directed motion of vesicles}
\begin{figure}[ht]
\centering
\includegraphics[width=\linewidth]{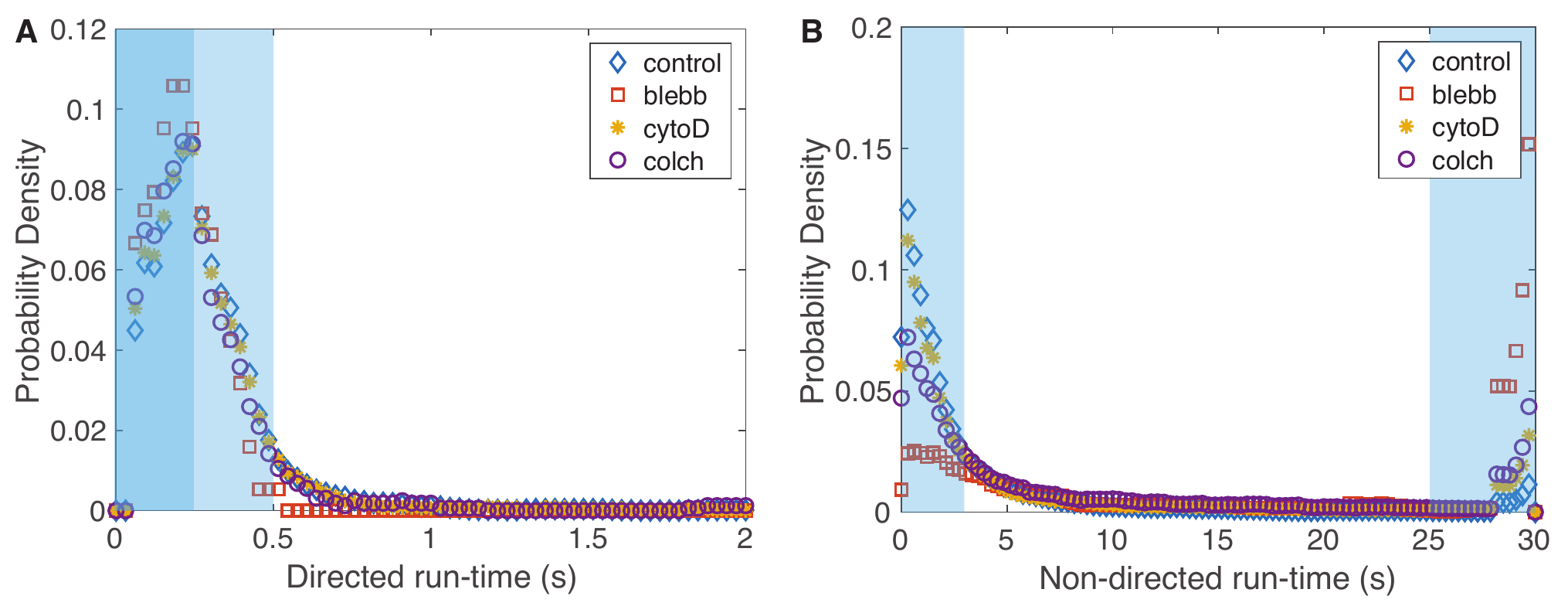}
\caption{\textbf{Directed run-times are robust, but non-directed run-times are sensitive to cytoskeletal perturbation.} (A) The directed run-time distribution indicates vesicles spend a similar amount of time undergoing directed motion independent of cytoskeletal perturbation.  These durations are relatively short, typically less than 500 ms. (B) Non-directed run-times indicate two timescales of random motion, short ($<$ 3 s) and long ($>$ 25 s) for all conditions. When cytoskeletal perturbations are introduced, there is a shift in non-directed motion from short to long run-times.  Inactivation of myosin-II has the strongest effect, shifting nearly half of vesicles to experience mainly non-directed motion.}
\label{fig:runtimes}
\end{figure}

In contrast to small-scale directed motion, non-directed motion of vesicles in the cytoplasm is not robust and does appear to be affected by drug treatments.  We identified two distinct groups for the run-time duration of non-directed motion: short duration ($\sim3$ seconds) and long duration ($\sim$25 seconds) (Fig. \ref{fig:runtimes}B).  A short duration non-directed run-time indicates a vesicle spends only a short amount of time moving randomly before transitioning to directed movement.  A long duration non-directed run time indicates a vesicle undergoes random motion for the majority of the measurement period.  To quantify the statistics, we integrate the probability density functions.  In control cells, 73\% of vesicles exhibit short run-times of non-directed motion (blue diamonds) and only 4\% exhibit long run-times.  However, when myosin-II is inhibited (Fig. \ref{fig:runtimes}B, red squares), there is a significant shift towards more vesicles undergoing non-directed motion for longer run-times: only 24\% display short run-times while 47\% display long run-times.  Thus, we see a shift from 4\% (control) to 47\% (blebbitstatin) of vesicles exhibiting non-directed run-times when myosin-II activity is inhibited.   Treatments affecting the cytoskeleton have an intermediate effect.  In cyto-D and colchicine treated cells 65\% and 50\% of vesicles, respectively, experience short run-times and 11\% and 15\% experience long run-times of random small-scale motion (Table \ref{tab:Non-directed_runtimes}).  These results suggest that in the unperturbed state (control), the majority of vesicles spend only a short amount of time undergoing non-directed motion before transitioning to directed motion.  When myosin-II is inhibited, this behavior shifts so that a large proportion of vesicles undergo non-directed motion for the majority of the measurement time.

\begin{table}[ht]
\centering
\begin{tabular}{|l|l|l|l|}
\hline
\textbf{Substrate Stiffness} & \textbf{Control} (nm/s) & \textbf{Cyto-D} (nm/s) & \textbf{Colchicine} (nm/s) \\
\hline
10 kPa & 525$\pm$385 & 595$\pm$674 & 456$\pm$288 \\
\hline
40 kPa & 539$\pm$370 & 585$\pm$507 & 562$\pm$385 \\
\hline
Glass & 555$\pm$371 & 589$\pm$478 & 641$\pm$446 \\
\hline
\end{tabular}
\caption{\label{tab:example} \textbf{Average velocities of vesicles undergoing directed motion.} (mean $\pm$ stdev).  Velocity distributions of directed motion are broad (see Fig. \ref{fig:VACF}A) and are not strongly affected by cytoskeletal perturbation.  Note that blebbistatin data is not available, due to the low occurrences of directed motion. }
\label{tab:Mean_velocities}
\end{table}

\begin{table}[ht]
\centering
\begin{tabular}{|l|l|l|l|l|}
\hline
\textbf{Time} (ms) & \textbf{Control} & \textbf{Cyto-D} & \textbf{Colchicine} & \textbf{Blebbistatin}  \\
\hline
500 & 94.3\% & 93.9\% & 93.9\% & 100.0\% \\
\hline
250 & 54.8\% & 57.2\% & 61.7\% & 71.4\% \\
\hline
\end{tabular}
\caption{\label{tab:example} \textbf{Statistics of directed run-times within a certain interval time.}  Time intervals correspond to areas of integration shown in Fig. \ref{fig:runtimes}A.  Most directed run-times were of duration 500 ms or less. (data shown for 10kPa substrate)}
\label{tab:Directed_runtimes}
\end{table}

\begin{table}[ht]
\centering
\begin{tabular}{|l|l|l|l|l|}
\hline
\textbf{Time} (s) & \textbf{Control} & \textbf{Cyto-D} & \textbf{Colchicine} & \textbf{Blebbistatin} \\
\hline
$<$ 3 & 73.4\% & 65.5\% & 50.2\% & 24.4\% \\
\hline
$>$ 25 & 4.0\% & 11.1\% & 15.1\% & 47.0\% \\
\hline
\end{tabular}
\caption{\label{tab:example}\textbf{Statistics of non-directed run-times within a certain interval time.} Time intervals correspond to areas of integration shown in Fig. \ref{fig:runtimes}B. In control cells, the majority of non-directed motions lasted less than 3 s before switching to directed motion.  When cytoskeletal perturbations are introduced, the distribution shifts indicating vesicles spend longer durations in non-directed states.  This effect is particularly strong for blebbistatin.}
\label{tab:Non-directed_runtimes}
\end{table}

\subsection*{Small-scale vesicle motion in cells on glass substrates is not strongly affected by drug treatments}
For cells cultured on a glass substrate, the vesicles in treated and untreated cells exhibit similar small-scale motion as shown by the MSD (Fig. \ref{fig:MSD_drugs}A).  This result is interesting because it suggests that perturbation of actin, myosin-II, or microtubules does not induce a strong effect on small-scale fluctuating vesicle motion for cells cultured on glass surfaces.  Any change in vesicle motion is significantly greater on softer substrates. 

\subsection*{On softer substrates, depolymerization of F-actin increases vesicle motion and depolymerization of microtubules decreases vesicle motion}
\begin{figure}[ht]
\centering
\includegraphics[width=\linewidth]{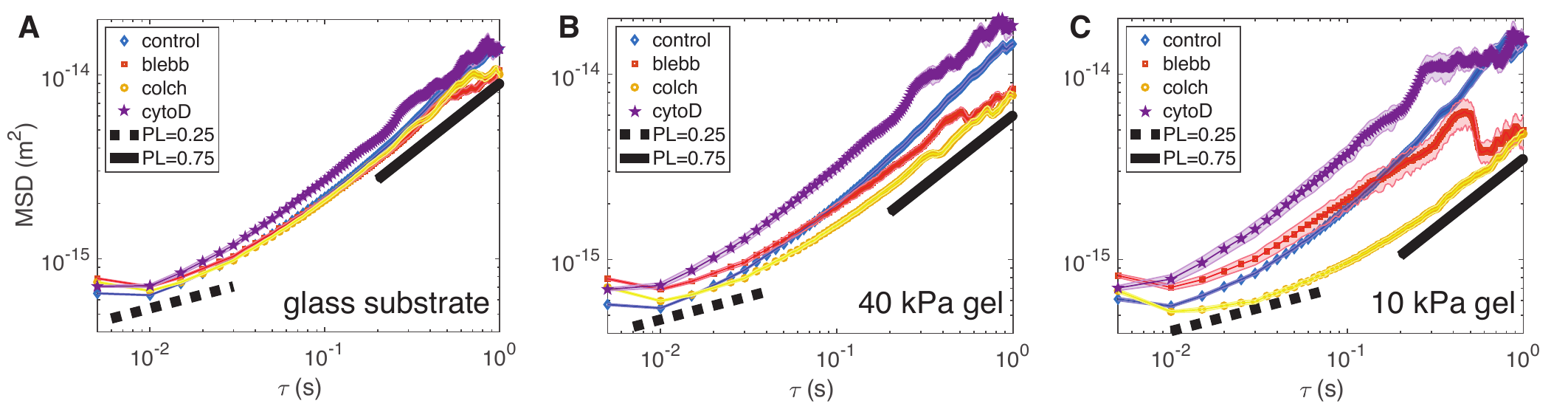}
\caption{\textbf{Vesicle motion is more sensitive to cytoskeletal perturbation on softer substrates.} (A) Vesicle dynamics on glass is not strongly affected by cytoskeletal perturbation.  However, the effect becomes more pronounced on (B) 40 kPA gels and (C) 10kPa gels.  Relative to the control vesicle motion, MSD plots suggest colchicine treatment decreases motion and cyto-D increases motion.(Number of experiments/cells/trajectories - glass: control 13/31/6738, blebb 8/19/4498, colch 8/16/2111, cytoD 8/18/3984; 40 kPa: control 10/30/3484, blebb 8/16/2412, colch 8/16/1324, cytoD 8/16/2139; 10 kPa: control 10/27/1855, blebb 8/15/773, colch 8/10/532, cytoD 8/12/797; S.E.M. shown as shaded background)}
\label{fig:MSD_drugs}
\end{figure}

In contrast to cells on glass, vesicle motion in cells on softer substrates is affected by cytoskeletal perturbations.  This effect increases with decreasing stiffness of the substrate (Fig. \ref{fig:MSD_drugs}B,C).  On softer substrates, depolymerizing microtubules leads to a decrease in vesicle mobility (yellow circles) and depolymerizing actin leads to an increase in vesicle mobility (purple stars).  The observed effect is stronger on 10 kPa gels (Fig. \ref{fig:MSD_drugs}C) relative to 40 kPa (Fig. \ref{fig:MSD_drugs}B) and is most evident at longer timescales.  This result aligns with the idea that microtubules function as highways for vesicle transport and that actin may serve to cage vesicles \cite{giner2007vesicle,Katrukha_Probing_2017}.


 \section*{Discussion}

Substrate stiffness is known to affect cell mechanics, thus it was expected to influence vesicle motion.  However, the MSD analysis (Fig. \ref{fig:MSD_stiffness}) and effective diffusion coefficients (Table \ref{tab:Diffusion_coefficients}) show substrate stiffness does not strongly effect vesicle motion.  A recent study of size and speed-dependent vesicle motion in cells \cite{Hu_Size_2017} suggests that in this regime of vesicle velocity ($v/r \sim 1$) vesicle motion due to an applied force is independent of vesicle size.  Their finding implies that in this regime, the mechanical properties of the cytoplasm may not have a strong effect on vesicle motion.  Our results provide further evidence supporting the study by Hu et al \cite{Hu_Size_2017}, since we do not observe a change in vesicle motion with substrate stiffness (which is expected to change cell mechanical properties).  While vesicle dynamics in healthy cells seems unaffected by substrate stiffness, the same cannot be said for cells treated with cytoskeletal perturbing agents.


For instance, when myosin-II activity is inhibited the vesicle diffusivity decreases and this affect is stronger of softer substrates. The VACF is a convenient tool to quantify the persistent dynamics of a system, for example, whether it behaves more like a solid, liquid, or gas \cite{boon1980molecular,Ahmed_SciRep_2015}.  When blebbistatin is present, the VACF exhibits the strongest initial anti-correlation of all treatments (Fig. \ref{fig:VACF}).  One interpretation is that the cytoplasm exhibits more solid-like behavior when myosin-II activity is inhibited.  Or in other words, myosin-II activity helps maintain a soft cytoplasmic environment.  Myosin-II is known to form myosin heavy chain isoforms which attach to actin filaments and pull them toward each other \cite{Murrell_Forcing_2015}.  Myosin-II activity may fluidize the cytoplasm (softening the environment) allowing vesicles to diffuse more freely, as supported by experimental and theoretical studies \cite{humphrey2002active,oriola2017fluidization}.  Since these motors can be activated locally, this is a possible mechanism for cells to tune their local cytoplasmic environment to facilitate or impede vesicle motion.  A similar effect of myosin-II inhibition has been observed in amoeba \cite{Reverey_Superdiffusion_2015}.

While the VACF was used in this study to quantify the persistent dynamics and calculate $D_\mathrm{eff}$, plots of MSD quantify the overall vesicle motion and suggest that actin filaments may impede local motion.  MSD analysis suggests that on softer substrates, vesicle motion increases in the absence of actin filaments as is visible by the increased amplitude of motion for cyto-D treated cells (Fig. \ref{fig:MSD_drugs}B,C purple stars).  Note that while this result is clear in the MSD, it  is not reflected in the $D_\mathrm{eff}$ obtained from integrating the VACF (Table \ref{tab:Diffusion_coefficients}) because this calculation is dominated by the initial anticorrelation (Fig. \ref{fig:VACF}B).  The notion of actin filaments acting as a barrier impeding motion is consistent with previous results where F-actin was found to trap non-functionalized quantum dots in the actin meshwork of COS-7 cells \cite{Katrukha_Probing_2017}.  

In addition to the VACF and MSD, the run-times of directed and non-directed motion shed light on the cytoskeleton's influence on vesicle motion.  It was observed that small-scale directed motion for vesicles in all treated and untreated cells lasted only about 500 ms (Fig. \ref{fig:runtimes}A), and velocities of these vesicles were not significantly changed by any drug treatment (Fig. \ref{fig:VACF}A and Table \ref{tab:Mean_velocities}).  Since these directed motions may not be driven by cargo transport motors (as they are not influenced by depolymerization of actin or microtubules), this robustness may reflect an underlying structure of the cytoskeletal network.  For instance, the mean directed run-time ($\sim 200$ ms) and the mean directed velocity ($\sim 500$ nm/s) provides a characteristic length of $\sim 100$ nm of the directed motion.  That value is on-par with the expected mesh size of the cytoskeleton in cells  and in this case may be related to the density of intermediate filaments \cite{sivaramakrishnan2008micromechanical,schopferer2009desmin}, because we observe the same behavior in the absence of actin filaments or microtubules.

As aforementioned, a recent study demonstrated that quantum dots ($r_{QD}= 15$ nm) in the cytoplasm of fibroblast-like cells (COS-7) can undergo either fast or slow diffusion \cite{Katrukha_Probing_2017}, and fast diffusion occurs when the actin is destabilized.  Our recorded measurements of fluctuating vesicles ($r_{ves}\sim 600$ nm) are in agreement with the slow diffusing quantum dots trapped in the actin meshwork.  In our case, the diffusion coefficient of vesicles was $D_{ves} = 1.5 \times 10^{-3}$ $\mu$m$^2/$s whereas for quantum dots, $D_{QD} = 0.6 \times 10^{-1}$ $\mu$m$^2/$s \cite{Katrukha_Probing_2017}.  This difference, $D_{QD}/D_{ves} \sim 40$, can be accounted for by the difference in radii of our objects, $r_{ves}/r_{QD} \sim 40$.  Similarly, a separate study measured the diffusion of peroxisomes ($r_{PO}\sim 200$ nm) in COS-7 cells to be $D_{PO} =6 \times 10^{-3}$ $\mu$m$^2/$s \cite{Lin_Active_2016}, where that difference can also be accounted for by the radii, $r_{ves}/r_{PO} \sim 3$.  The results we found are also remarkably similar to measurements of carbon nanotube diffusion in COS-7 cells ($l_{CNT}\sim 100-300$ nm), including the MSD and the velocity distribution \cite{Fakhri_High_2014}.  Together, our measurements of small fluctuating motion of vesicles,  combined with studies of quantum dots \cite{Katrukha_Probing_2017}, peroxisomes \cite{Lin_Active_2016}, and carbon nanotubes \cite{Fakhri_High_2014}, suggest the local diffusion in the fibroblast cytoplasm may be similar for objects varying in size from tens of nanometers to micron-sized.  While this result may be expected in a simple viscous fluid, how this occurs in a heterogeneous complex viscoelastic medium is not clear.  Further studies are necessary to probe why seemingly diffusive-like behavior is repeatedly observed in these complex nonequilibrium systems. Future studies would also benefit from specific labeling of vesicles since analyzing a non-homogeneous, undefined set of vesicular cargoes limits interpretation of data.  The vesicles tracked in this study are of mixed identity (e.g. peroxisomes, early endosomes, late endosomes, lysosomes, autophagosomes) and each could have distinct motility behavior.

\section*{Conclusion}
In this study, we found that small-scale displacement fluctuations of vesicles in fibroblasts do not depend on their substrate stiffness.  However, drugs perturbing the cytoskeleton (actin, microtubules, and myosin-II) do affect vesicle dynamics and this affect is increased on softer substrates.  It was found that myosin-II has the largest effect on vesicle motion, as it decreases the diffusion coefficient of the local cytoplasm.  Actin and microtubule filaments also modulate the mobility of vesicles but to a lesser extent.  Our results support the idea that actin may trap vesicles and impede local movement \cite{Katrukha_Probing_2017}.  This increased understanding of the role of the cytoskeleton in local vesicle motion may help inform developments for targeted drug therapies in the future.

\section*{Materials and Methods}

\subsection*{Hydrogel fabrication and cell culture}
Polyacrylamide (PA) gels of varying elastic moduli ($E$ = 10 and 40 kPa) were fabricated as previously reported \cite{Knoll_Contractile_2015}.  Gel stiffness was modulated by mixing acrylamide and bisacrylamide according to specifications reported in a well-established protocol \cite{Kandow_Methods_2007}.  Monkey kidney fibroblast cells (CRL-1651, ATCC, Manassas, VA) were plated sparsely on PA gels (2500 cells/cm$^2$) and immersed in media comprised of Dulbecco’s Modified Eagle’s Medium (Corning, Corning, NY) (4.5 g/L glucose, 4 mM L-glutamine) supplemented with 10\% fetal bovine serum (Sigma-Aldrich, St. Louis, MO) and 1\% Penicillin Streptomycin (Corning, Corning, NY).

\subsection*{Cytoskeletal purturbation}
Three agents were used to perturb the cytoskeleton: cytochalasin-D, colchicine, and blebbistatin.  Cytochalasin-D (5 $\mu$M) was used to depolymerize F-actin filaments \cite{goddette1986actin,cooper1987effects}.  Colchicine (10 $\mu$M) was used to depolymerize microtubules \cite{dalbeth2014mechanism,kemkemer2000nematic}.  Blebbistatin (25 $\mu$M) was used to inactivate myosin-II motors \cite{Kovcs_Mechanism_2004}.  In all cases cells were incubated in the perturbing agent for at least one hour before, and throughout the imaging process.  After imaging, the perturbing agent was washed out and replaced with fresh cell media to observe cell recovery and viability.

\subsection*{Microscopy}
All experiments were performed using differential interference microscopy on an Olympus IX81 microscope with a 40X UApo N340 water immersion objective (NA 1.15) (Olympus America Inc., Center Valley, PA) mounted on a vibration isolation table (Newport Corporation, Irvine, CA).  An environmental chamber enclosed the experimental platform, and maintained cell culture conditions throughout imaging (5\% $CO_{2}$, 70\% humidity and 37$^{\circ}$C).  Images were acquired  with a Neo sCMOS camera (active pixels 1392 x 1040, resolution of 165 nm per pixel) (Andor Technology, Belfast, Northern Ireland) at 200 frames per second for a total duration of 30 seconds.  

\subsection*{Image analysis and processing}
Image stack alignment to remove drift was completed with a subpixel registration algorithm \cite{thevenaz1998pyramid}.  Vesicle motion was tracked using an algorithm for precise particle tracking by polynomial fitting with Gaussian weight \cite{rogers2007precise} in MATLAB (The MathWorks, Natick, MA, USA).  This algorithm was developed specifically for tracking endogenous vesicles in living cells with noisy backgrounds at nanometer precision. In this study, we applied the algorithm to differential interference contrast microscopy images (representative image shown in Figure S2).  To compute the precision of our vesicle tracking analysis, we calculate the error to be $\sqrt[]{\mathrm{MSD}(\tau = \mathrm{5 \, ms}}) \sim 25$ nm \cite{rogers2007precise}.  This tracking error is consistent with that measured in other comparable studies \cite{Fakhri_High_2014,Katrukha_Probing_2017,guo2014probing}.  Note that any trajectory of a vesicle that moved within 3 $\mu$m of the cell boundary or nucleus was discarded.  Similarly, trajectories of neighboring vesicles that came within 500 nm were discarded.  The radii of vesicles were calculated, as in Rogers et al \cite{rogers2007precise}, and found to be $r = 619 \pm 95$ nm.  Vesicle motion was not observed to depend on vesicle size, in the small range of vesicle sizes observed ($400 < r < 800$ nm).  All computations, statistical analysis, and plots were generated in MATLAB.

\subsection*{Temporal mean squared displacement analysis}
To discriminate between directed and non-directed motion, vesicle dynamics were analyzed using a temporal Mean Squared Displacement ($t$MSD) algorithm to extract the local dynamics of the vesicle as a function of time as done previously \cite{Ahmed_Measuring_2012}.  Briefly, the algorithm estimates the local behavior for time, $t$, by calculating the $t$MSD of a rolling window centered about the time of interest,
\begin{equation}
t\mathrm{MSD}(\tau)=\langle\vert\mathbf{r}(t^{\prime}+\tau)-\mathbf{r}(t^{\prime})\vert^{2}\rangle_{(t-t_{\mathrm{max}}/2<t^{\prime}<t+t_{\mathrm{max}}/2)}
\end{equation}
where $<...>$ indicates an average over timescale $\tau$, $\mathbf{r}(t)$ is the position vector as a function of time, $t^\prime$ is the time in the rolling window, $\tau$ is the timescale, and $\tau_{\mathrm{max}}$ is the width of the rolling window ($\tau_{\mathrm{max}} = 500$ ms).  The $t\mathrm{MSD}$ is then fitted to a power law of the form, $t\mathrm{MSD}(t) = C\tau^{\alpha}$ on the interval 100–160 ms, and the power law scaling, $\alpha(t)$, can be extracted as a function of time to indicate the type of diffusion the vesicle is undergoing ($\alpha < 1$, sub-diffusive; $\alpha = 1$, Brownian; $\alpha > 1$, super-diffusive).  Based on previous calibrations, $\alpha_{\mathrm{directed}} \geq 1.4$ was chosen as a conservative threshold for directed motion and $\alpha_{\mathrm{non-directed}} < 1.4$ was considered non-directed.  Only trajectories of particles longer than 200 frames were kept for analysis. An example is shown in Figure S1.

\subsection*{Correlation functions}
In addition to the $t$MSD we used the velocity autocorrelation function (VACF) and the van Hove correlation function (VHC) to quantify dynamics.  The VACF, 
\begin{equation}
\psi(\tau)=\frac{\langle \textbf{v}(t)\cdot \textbf{v}(t+\tau)\rangle}{\langle[v(t)]^{2}\rangle}
\end{equation}
quantifies persistence of directional motion by its characteristic decay with $\tau$.  The effective diffusion coefficient of the vesicle was calculated from the VACF using the Green-Kubo relation, $D_{\mathrm{eff}}=v_{0}^2\int_{0}^{\infty}\psi(\tau)d\tau$ \cite{boon1980molecular}.  We use this integral approach to calculate $D_{\mathrm{eff}}$ instead of fitting $\mathrm{MSD}=4D\tau + v^2\tau^2$ because we often observe our vesicle motion to be sub-diffusive.  The VHC, quantifies the probability distribution of vesicle displacements, $P(\Delta x,\tau)$ where $\Delta x=x(t+\tau)-x(t)$ for a given timescale, $\tau$.  The overall shape of the VHCs provide valuable information on the vesicle fluctuations \cite{Ahmed_SciRep_2015}.

\subsection*{Data availability}
The datasets generated during and/or analysed during the current study are available from the corresponding author on reasonable request.

\bibliography{sample}

\section*{Acknowledgements}
The authors acknowledge fruitful conversations with L. Kapitein (Utrecht University), S. Arumugam (University of New South Wales), and E. Fodor (University of Cambridge).  WWA, DP, and PB-P acknowledge financial support from the Cal State Fullerton Research Scholarly and Creative Activity Grant, PB-P acknowledges the Dan Black Physics Scholarship, SKK acknowledges the UIUC ECE Fellowship, and WA and TS acknowledge NSF CMMI 13-00808.

\section*{Author contributions statement}
D.P., P.B-P., W.A. analyzed the data and wrote the manuscript. S.K. and W.A. designed and conducted all experiments.  T.S. and W.A. supervised the project.  All authors reviewed the manuscript.

\section*{Author competing interests statement}
The authors declare no competing interests.

\end{document}